# Scanning-free imaging through a single fiber by random spatio-spectral encoding

Sylwia M. Maliszewska[1,3†], Ori Katz[2,3†*], Mathias Fink[2], Sylvain Gigan[3]

[1]Institute of Physics, Faculty of Physics, Astronomy and Informatics, Nicolaus Copernicus University, Grudziadzka 5, 87-100 Torun, Poland

[2] Institut Langevin, UMR7587 ESPCI ParisTech and CNRS, INSERM ERL U979, 1 Rue Jussieu, 75005 Paris, France

[3]Laboratoire Kastler Brossel, UMR8552 of CNRS and Université Pierre et Marie Curie, 24 rue Lhomond, 75005 Paris, France

*Corresponding author:ori.katz@espci.fr

We present an approach for two-dimensional (2D) imaging through a single single-mode or multimode fiber without the need for scanners. A random scattering medium placed next to the distal end of the fiber is used to encode the collected light from every imaged pixel with a different random spectral signature. 2D objects illuminated by a white-light source are then imaged from a single measured spectrum at the fiber's proximal end. The technique is insensitive to fiber bending, an advantage for endoscopic applications.

Optical endoscopes emerged as a very efficient and robust tool for imaging the interior of the human body in a minimally invasive manner. Recent efforts have been focused on advancing towards micro-endoscopy, improving the imaging resolution, field of view (FOV) and probe miniaturization by employing a variety of imaging approaches and designs [1]. Several common configurations utilize fiber bundles constructed from thousands of individual cores, each carrying one image pixel information. Unfortunately, fiber bundles suffer from a limited imaging resolution and low fill factors dictated by the individual fiber's core and cladding diameters, respectively. Alternatively, the image information can be delivered by a single multimode fiber (MMF), where each transverse fiber mode transmits a different pixel element [2, 3]. The inherent problem of phase randomization and mode mixing in propagation through a MMF can be compensated for by pre-measuring the complex input-output transmission matrix (TM) for the propagating fiber modes [3], or by phase conjugation [2, 3], effectively unmixing back the propagation effects. Recently, impressive results have been obtained by applying the principles of these approaches using digital cameras and computer-controlled spatial light modulators (SLMs) [4-9]. These approaches have also been applied to overcome the spatial resolution limit of fiber bundles [10, 11]. However, a major drawback of these TM-based approaches is their sensitivity to fiber bending. Any act of bending results in an unpredictable change in mode mixing and thus a variation in the fiber's TM, resulting in image distortion.

It is possible to use a single single-mode fiber (SMF) as a bend-insensitive endoscope, but in order to obtain 2D images, a mechanical scanning head should be mounted at the distal end of a fiber, sacrificing frame rate and probe size [1]. The requirement for 2D mechanical scanning can be reduced to one dimension (1D) by spectrally dispersing the light in the other dimension by a dispersive component at the distal end [12]. The spectrally-encoded line of pixels (where each pixel is represented by a single wavelength) can be detected with a line-spectrometer, but still requires 1D mechanical scanning for 2D imaging. Recently, 1D spectral encoding was combined with a fiber bundle to produce 2D images without mechanical scanning [13], and 2D spectral encoding was demonstrated by constructing an even more complex 2D spectral disperser [14]. However, one of the major hurdles in such spectral encoding techniques is the difficulty in the design and high-precision fabrication of the miniature dispersive element.

Here, we present a simple 2D spectral encoding approach that is based on scattering from a random medium placed at the distal end of a single fiber. We show that even a simple diffuser can serve as a random 2D spectral dispersing element. As random scattering, even from a thin diffuser, is spectrally dependent [15, 16], the light scattered from a diffuser is inherently encoded with different random spectral signatures for each spatial pixel position (Fig.1a). This random, but fixed, encoding can be measured in a simple calibration procedure (Fig.1a). Given the spatio-spectral encoding, the image of an object placed next to the distal end and illuminated by natural light can be reconstructed from the spectrum measured at the proximal end (Fig.1b), using a linear algebra based reconstruction [17], or a compressive sensing (CS) reconstruction algorithm [18].

The concept is presented in Fig.1. The simplest implementation is comprised of a SMF with a spectrometer at its proximal end and a random scattering medium placed next to its distal end. The measurement process involves two stages: The first is a 'calibration' procedure, mapping the scattering sample, where the different encoded 'speckled' spectra, $S_x(\lambda)$, for each imaging pixel position, $x$, are acquired. The spectra are acquired sequentially by scanning the position of a broadband luminous point source over each pixel position (Fig.1a). The set of measured spectra constructs a spatio-spectral mapping matrix $H_{\lambda,x} = S_x(\lambda)$ that contains the spectral fingerprints of each image pixel. The second step is the single-shot imaging step: the recording of a spectrum when a 2D object is placed at the imaging plane and is incoherently illuminated by a broadband source, such as natural light, with the same spectrum as the one used in the calibration step (Fig.1b). As an extended object is essentially comprised of a set of luminous point sources, the measured spectrum in the imaging step, $I_\lambda^{meas}$, is simply the intensity sum of the corresponding pre-calibrated spectra for all luminous points of the object:

$$I_\lambda^{meas} = \sum_x H_{\lambda,x} I_x^{obj} \qquad (1)$$

where $I_x^{obj}$ is the achromatic object's 2D transmission (or reflection) pattern (intensity transmission at each spatial position $x$). Finally the image of the object, $I_x^{obj}$, is reconstructed in a strictly algorithmic step.

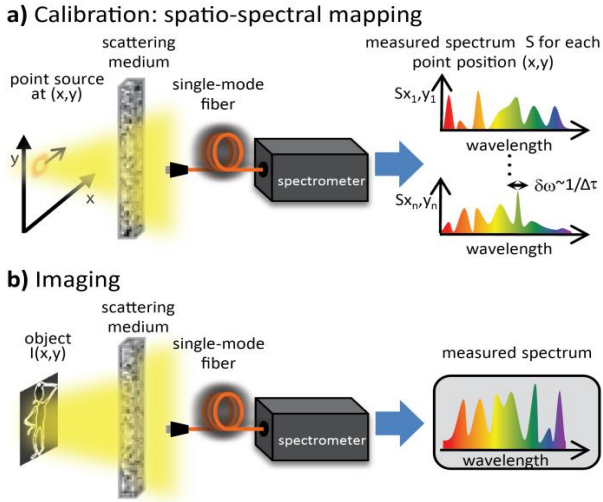
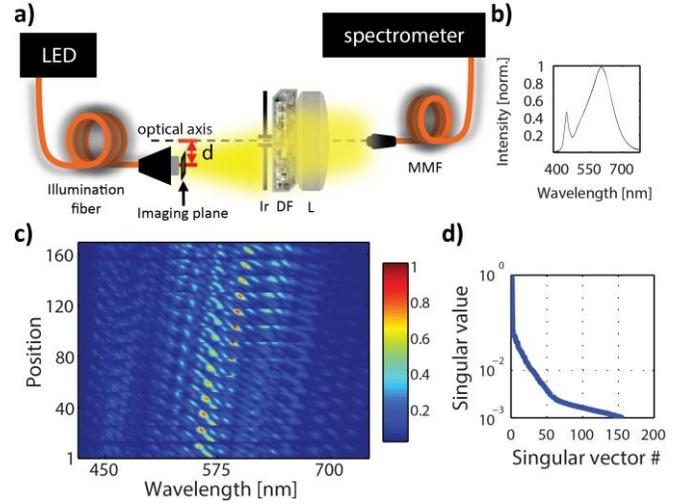

Fig. 1. The random spectral encoding concept is comprised of two steps: a) Calibration of the scattering sample: a spectrally broadband point source is scanned over the different imaged pixels. For each pixel position, $x$, the randomly encoded spectrum that is coupled to the fiber $S_x(\lambda)$ is recorded, constructing a spatio-spectral mapping matrix $H_{\lambda,x}$. b) Single-shot imaging: a spectrum that is measured when an incoherently illuminated object is placed at the imaging plane, $I_\lambda^{meas} = \sum_x H_{\lambda,x} \cdot I_x^{obj}$, contains sufficient information to reconstruct the object's 2D intensity pattern, $I_x^{obj}$, with the knowledge of $H_{\lambda,x}$.

Fig. 2. Experimental spatio-spectral encoding measurement. a) Setup: A white light fiber-coupled LED is used as the light source. The illumination MMF is positioned at a distance $d$ from the optical axis. The spatio-spectral encoding matrix is acquired by scanning a pinhole across the 400µm illumination fiber core, serving as the imaging plane. The light from the pinhole illuminates a diffuser (DF), and is coupled by a lens (L) to the collection fiber, connected to a spectrometer, An iris (Ir) is used to assure that only a single spatial mode (speckle grain) is coupled to the detection MMF. b) The LED spectrum, c) a measured spatio-spectral mapping matrix, $H_{\lambda,x}$, i.e. the spectra for different positions of the pinhole (see Media 2 for spectral spatial projection). d) Singular value decomposition, SVD, of $H_{\lambda,x}$ used to estimate the number of independent spectral channels (see text).

The reason that a random scattering medium acts as a random spectral encoder is that at each illumination wavelength the light scattered from the medium produces a different random speckle pattern [19]. As the speckle patterns differ for each wavelength, measuring the spectrum at a single spatial speckle position causes the appearance of random spectral modulations in the detected spectrum [16], even for a thin diffuser [15]. The spectral features width, $\delta\omega$, is the 'speckle spectral decorrelation width' [19]. Similar to diffraction from a grating, $\delta\omega$ is inversely related to the scattered light path lengths distribution between the input point and measurement point, $\Delta l$: $\delta\omega \approx 2\pi c/\Delta l$, where $c$ is the speed of light. For a thin scatterer $\Delta l$ is simply geometrical, while for a thick multiply scattering sample $\Delta l$ depends on the scattering properties of the medium.

A proof-of-concept experimental realization of the approach is presented in Fig.2a. As a broadband illumination source we use the spatially-incoherent light from a fiber-coupled warm light LED (Thorlabs MWWHF1) providing a power of ~3mW and a bandwidth of ~140nm FWHM (Fig.2b) through a Ø 400-µm core MMF (0.39 NA). We used the plane just in front of the 400-µm fiber facet as the imaging plane. A Newport 5° light shaping diffuser placed 150mm from the imaging plane serves as the scattering medium. An f=75mm lens is placed behind the diffuser and a MMF (Ø 50-µm core, 0.22 NA) placed 150mm from the lens couples the scattered light to a spectrometer (Andor Shamrock SR-303i with a 300 l/mm grating and Andor iXon EMCCD camera). To maximize the spectral modulations contrast, an iris in front of the diffuser is set to match the speckle size at the collection fiber facet plane to the collection fiber 50-µm core, so that only a single spatial mode (speckle) spectrum is measured with maximal intensity (iris diameter ~1mm). As the entire spectrum of the coupled light is measured by the spectrometer, the collection fiber simply serves as a 'light tube', and the measured spectrum is insensitive to the fiber bending. To obtain a sufficiently narrow spectral feature width, $\delta\omega$, (see below) the imaged area is positioned 7mm from the optical axis, where the scattered light's optical path lengths distribution is sufficiently broad [15].

To acquire the spatio-spectral mapping matrix, a 50-µm pinhole placed in front of the illuminating fiber serves as a point source. Its position is scanned over a 300x300 µm² area in 25-µm steps (13x13 points in total), and the collected spectrum for each position is recorded with a 4s exposure time. An example for an experimentally measured mapping matrix $H_{\lambda,x}$ is presented in Fig.2c (rotated 90 degrees for viewing purposes). Although no special effort to stabilize the system was made, the spatio-spectral encoding was stable for more than 13 hours, and was insensitive to fiber bending as expected (>0.99 spectral correlation). We then used this measured $H_{\lambda,x}$ to reconstruct several objects: a moving single point (Fig.3) and digits from the 2nd group of USAF 1951 Test Target (Fig.4).

As equation (1) is expressed in matricial form by $I^{meas} = HI^{obj}$, the reconstruction problem is a linear inverse problem where the unknown vector $I^{obj}$ (comprised of $N$ pixels) needs to be found from a set of $M$ equations of the spectral intensities values. In the ideal noiseless case where $M = N$ the solution is given by inversion $I^{obj} = H^{-1}I^{meas}$. However, in the presence of noise, the 'best fit' least-squares solution is given by the generalized inverse (the 'pseudoinverse') $H^+$, rather than $H^{-1}$, such as the Tikhonov regularized inversion [15]:

$$I^{obj} = H^+ I^{meas} \qquad (2)$$

(note that $H^+$ is not the conjugate transpose $H^\dagger$).

As a first demonstration we used the pinhole from the calibration step placed at different positions on the mapped 300x300 µm² area. Figure 3 presents the results of pseudoinverse reconstruction for three different pinhole positions together with the corresponding measured spectra, $I_\lambda^{meas}$, captured with an exposure time of 2s. A video of the reconstruction of the moving pinhole is given in Media 1. The object is reconstructed using the Moore-Penrose matrix inverse (implemented in Matlab by "\"), which is based on a singular value decomposition (SVD) of the calibration matrix $H$ with a manually set tolerance (threshold). We set the threshold to select only the 49 highest singular values of the matrix, as smaller singular values mostly originate from measurement noise (Fig.2d), and are better set to be treated as zero. Setting the tolerance to lower values does not contribute or reduces the performance of the pseudoinverse reconstruction.

As the reconstruction problem is a linear one, to uniquely reconstruct an $N$–pixel object, $K > N$ linearly independent equations (projections) should be available. In the noiseless case $K$ is given by the rank of the matrix $H$. In our spatio-spectral encoding case, $K$ is not the number of spectral bins of the spectrometer, $M$, as measurements at spectral shifts smaller than $\delta\omega$ are essentially identical, and correspond to similar spatial speckled projection patterns at the image plane (Media 2). The true number of independent spectral components (projections) can be estimated by $K \approx \Delta\omega/\delta\omega$, with $\Delta\omega$ being the spectral bandwidth of the illumination, yielding a value of ~300nm/5nm=60 in our experimental parameters. $K$ can be also estimated from the SVD of $H$ (Fig.2d), where indeed it seems that the highest ~60 singular values are significantly larger than the rest.

This inherent limitation of linear inversion can be overcome by exploiting a class of reconstruction algorithms known as Compressed Sensing (CS). Using these algorithms the conventional restriction of the Nyquist sampling rate can be surpassed if additional a-priori information on the object is available, as is the case with natural objects [18, 20]. The object can thus be reconstructed even from $M < N$ random sampled measurements of it, as long as some prior assumption on its, such as its sparsity in some transform domain, holds, as frequently used in image compression [20]. The result of applying a CS-based reconstruction on the experimental data of Fig.2 is presented in Fig.3. The CS reconstruction was implemented following the recipe given in [20]. One can see the reduced background noise and increased resolution [21] in the CS reconstruction compared to the pseudoinverse in this case of a very sparse object.

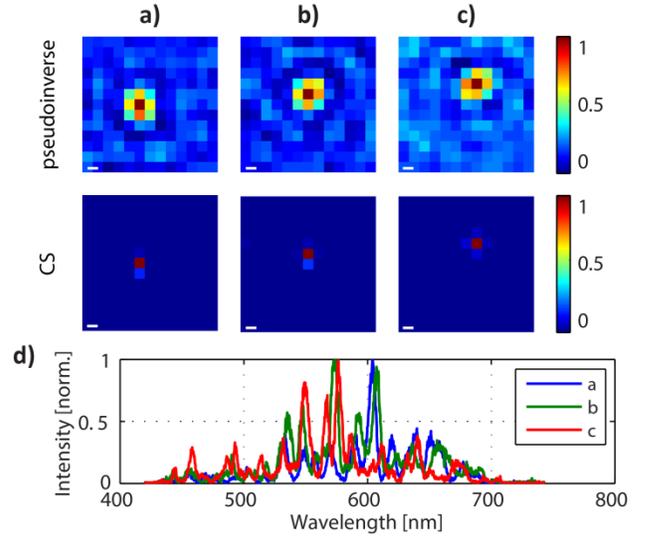

Fig. 3. (a-c) Experimental reconstruction of a moving point-source object at three points (see Media 1). top row: pseudoinverse reconstruction, bottom row: Compressive-Sensing (CS) reconstruction. (d) The raw measured spectra used for the reconstructions of (a-c). Scale bar: 25 µm.

Next we imaged more complex objects: the digits 1, 3 and 4 from the USAF 1951 Test Target's 2nd group. The results of these measurements are shown in Fig.4. In these experiments we mapped a 400x400 µm² area, successfully reconstructing these simple objects. The measured spectra in these experiments (Fig.4) exhibit a lower modulation contrast than the spectra measured with the point-object (Fig.3d); they all resemble the source spectrum (Fig.2b). This is an expected result, as each spectrum in Fig.4 is the sum of several spectra of the points constituting the object.

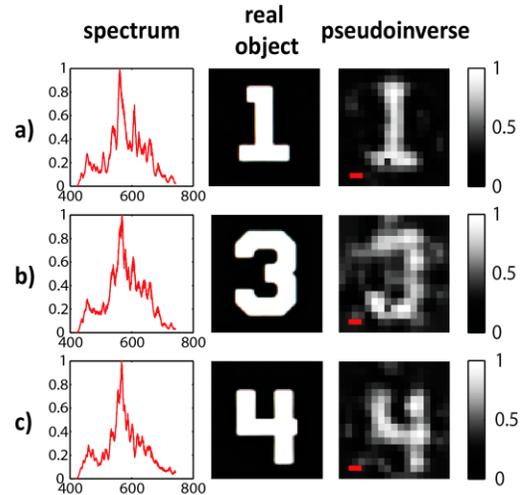

Fig. 4. (a-c) Pseudo-inverse reconstruction from a single measured spectrum of digits from USAF 1951 Test Target. The raw measured spectrum is plotted on the left column. Scale bar: 50 µm.

We have shown that single-shot 2D incoherent imaging can be performed through a single fiber by random spectral encoding. Our approach is intimately related to the recent use of scattering media for disorder-based spectrometers [22]. In fact, our experiments can be considered as the 'inverse' experiments, using similar spatio-spectral coupling of randomly scattered light. Utilizing a fixed stable scattering medium as the spectral encoder instead of relying on the knowledge of the MMF's TM our approach is insensitive to fiber bending. In addition, the use of spatially incoherent light potentially eliminates speckle imaging artifacts. Our approach is also related to single-antenna imaging by time-reversal [23], where spatio-temporal coupling by random scattering is exploited for imaging [23, 24]. However, while time-reversal techniques are inherently coherent, we use intensity only measurements for incoherent imaging.

Our method is bend-insensitive and produces two-dimensional images without any use of scanners. However, there are some important limitations that should be taken into account. An important practical limitation while using natural light is the inherent low light collection efficiency of the approach, resulting from the fact that only a single spatial mode (a single speckle) of the scattered light is collected. This may be overcome by using swept sources or by carefully designing the random medium [22]. Another limitation is the assumption of spectrally independent object's transmission or reflection, although the system can be calibrated in the case of known spectral constituents of the objects. An additional limitation is the requirement to be able to resolve the low contrast spectral modulations measured for complex objects; the contrast of the modulations decreases as the square root of the number of bright resolution cells of the object (summed-up speckled spectral patterns).

The choice of scattering medium cannot be random: a thick multiply scattering medium such as white paint [25] would be constructive in the sense of a smaller $\delta\omega$ and, as a consequence, a larger number of potential imaged resolution cells. However, thick scattering samples exhibit increased back-reflections and thus loss of signal compared to the near unity transmission of a simple diffuser.

The imaging resolution of the technique is inherently diffraction-limited, and set by the scattering angle of the random medium. This has been recently shown to surpass the resolution of the fiber NA [25]. However, as the number of independent spectral components limits the number of imaged resolution cells (with some multiplicative factor when CS reconstruction is used), increasing the resolution requires a decrease in FOV. The FOV could be extended by the use of bundles, where each core is 2D spectrally encoded, similar to the recent TM-based approach [11].

Extension to three-dimensional imaging is also possible as the scattered speckle pattern changes with axial position [26]. Future work would focus on the potential miniaturization of the random encoding apparatus and on combining the illumination and detection in a single fiber.

This work was funded by the European Research Council (grant 278025). O. K. is supported by the Marie Curie Intra-European Fellowship for career development. S.M. acknowledges support from European Social Fund's project POKL.04.01.01-00-081/10 WZROST

† These authors contributed equally to this work.

References:

1. G. Oh, E. Chung, and S. H. Yun, Opt. Fiber Technol. **19**, 760 (2013).
2. A. A. Friesem, U. Levy, and Y. Silberberg, Proc. IEEE **71**, 208 (1983).
3. A. Gover, C. Lee, and A. Yariv, JOSA **66**, 306 (1976).
4. R. Di Leonardo, and S. Bianchi, Opt. Express **19**, 247-254 (2011).
5. S. Bianchi, and R. Di Leonardo, Lab on a Chip **12**, 635-639 (2012).
6. T. Čižmár, and K. Dholakia, Nat. Commun. **3**, 1027 (2012).
7. I. N. Papadopoulos, S. Farahi, C. Moser, and D. Psaltis, Opt. Express **20**, 10583 (2012).
8. Y. Choi, C. Yoon, M. Kim, T. D. Yang, C. Fang-Yen, R. R. Dasari, K. J. Lee, and W. Choi, Phys. Rev. Lett. **109**, 203901 (2012).
9. G. Kim, and R. Menon, Appl. Phys. Lett. **105**, 061114 (2014).
10. E. R. Andresen, G. Bouwmans, S. Monneret, and H. Rigneault, Opt. Lett. **38**, 609 (2013).
11. D. Kim, J. Moon, M. Kim, T. D. Yang, J. Kim, E. Chung, and W. Choi, Opt. Lett. **39**, 1921 (2014).
12. D. Yelin, I. Rizvi, W. White, J. Motz, T. Hasan, B. Bouma, and G. Tearney, Nature **443**, 765 (2006).
13. N. Bedard, and T. S. Tkaczyk, J. Biomed. Opt. **17**, 0805081 (2012).
14. K. Goda, K. Tsia, and B. Jalali, Nature **458**, 1145 (2009).
15. E. Small, O. Katz, and Y. Silberberg, Opt. Express **20**, 5189 (2012).
16. E. Small, O. Katz, Y. Guan, and Y. Silberberg, Opt. Lett. **37**, 3429 (2012).
17. S. Popoff, G. Lerosey, M. Fink, A. C. Boccara, and S. Gigan, Nat. Commun. **1**, 81 (2010).
18. E. J. Candes, and M. B. Wakin, Signal Processing Magazine, IEEE **25**, 21 (2008).
19. I. Vellekoop, A. Lagendijk, and A. Mosk, Nat. Photonics **4**, 320 (2010).
20. O. Katz, Y. Bromberg, and Y. Silberberg, Appl. Phys. Lett. **95**, 131110 (2009).
21. S. Gazit, A. Szameit, Y. C. Eldar, and M. Segev, Optics Express **17**, 23920-23946 (2009).
22. B. Redding, S. F. Liew, R. Sarma, and H. Cao, Nature Photonics **7**, 746-751 (2013).
23. M. Fink, Sci. Am. **281**, 91 (1999).
24. F. Lemoult, M. Fink, and G. Lerosey, Nat. Commun. **3**, 889 (2012).
25. I. N. Papadopoulos, S. Farahi, C. Moser, and D. Psaltis, Opt. Lett. **38**, 2776 (2013).
26. Y. Bromberg, O. Katz, and Y. Silberberg, Phys. Rev. A **79**, 053840 (2009).